\documentclass{ifacconf}

\usepackage{graphicx}      % include this line if your document contains figures
\usepackage{natbib}        % required for bibliography
\usepackage{amsmath,amssymb,amsfonts}
\usepackage{algorithm,algpseudocode}
\usepackage{subfigure}
\usepackage{textcomp}
\usepackage{diagbox}
\usepackage{makecell}
\usepackage{multirow}
\usepackage{booktabs}
\usepackage{arydshln}
\usepackage{blkarray}
\usepackage{xcolor}

\newcommand{\E}{{\mathbb E}}
\newcommand{\Zbb}{\mathbb Z}
\newcommand{\Rbb}{\mathbb R}

\newcommand{\lm}{{\rm l}}
\newcommand{\mm}{{\rm m}}

\newcommand{\Ab}{\mathbf A}
\newcommand{\Bb}{\mathbf B}
\newcommand{\Cb}{\mathbf C}

\newcommand{\Yb}{\mathbf  Y}
\newcommand{\Zb}{\mathbf  Z}

\newcommand{\Ical}{\mathcal  I}
\newcommand{\Ccal}{\mathcal  C}

\newcommand{\Span}{\operatorname{span\,}}
\newcommand{\rank}{\operatorname{rank\,}}

\begin{document}
\begin{frontmatter}

\title{Modeling and Topology Estimation of Low Rank Dynamical Networks\thanksref{footnoteinfo}} 
% Title, preferably not more than 10 words.

\thanks[footnoteinfo]{This work is supported by the National Natural Science Foundation of China under Grant Nos. T2525017, 62533002, T2421004 and 62503020, 
the National Key Research and Development Program of China under Grant No. 2022YFA1008400, 
the Jiangsu Provincial Scientific Research Center of Applied Mathematics under Grant No. BK20233002,
the Beijing Natural Science Foundation under Grant No. 4254098,
and the Postdoctoral Fellowship Program of CPSF under Grant No. GZC20240041.}

\author[First]{Wenqi Cao} 
\author[Second]{Aming Li} 

\address[First]{Center for Systems and Control, School of Advanced Manufacturing and Robotics, Peking University, Beijing, China 
    (e-mail: wenqicao@pku.edu.cn).}
\address[Second]{Center for Systems and Control, School of Advanced Manufacturing and Robotics, and Center for Multi-Agent Research, Institute for Artificial Intelligence, Peking University, Beijing, China
    (e-mail: amingli@pku.edu.cn).}

\begin{abstract}                % Abstract of 50--100 words
Conventional topology learning methods for dynamical networks become inapplicable to processes exhibiting low-rank characteristics.
To address this, we propose the low rank dynamical network model which ensures identifiability.
By employing causal Wiener filtering,
we establish a necessary and sufficient condition that links the sparsity pattern of the filter to conditional Granger causality. 
Building on this theoretical result, we develop a consistent method for estimating all network edges.
Simulation results demonstrate the parsimony of the proposed framework and consistency of the topology estimation approach.
\end{abstract}

\end{frontmatter}
%===============================================================================

\section{Introduction}

Graphical models and dynamical networks provide a powerful framework for characterizing interaction patterns in multivariate systems, 
with broad applications across fields such as economics, 
ecological evolution and sociology. 
As system dimensionality increases, 
the corresponding graph topology often exhibits sparsity, 
implying each unit interacts directly with only a limited number of neighbors. 
This structural property not only provides critical functional insights 
but also renders the estimation of graph topology from data essential to systems identification and control. See, for example, \cite{Salapaka12tac}, \cite{Lin13}, \cite{Materassi21}, \cite{giap25survery} and \cite{Zorzi25tac}.\\
In the topology learning and parameter identification of dynamical networks, the invertibility of the spectral density is often a sufficient condition for a well-posed identification problem and a consistent estimate.
In addition, an inverse spectral condition %w.r.t. the conditional dependency 
is commonly employed to determine the edges between nodes, i.e.,  
\begin{align}\label{eq:cond_invPhi}
 [\Phi(e^{i\theta})^{-1}]_{kh}=0 \Leftrightarrow \Yb_{\{k\}} \perp \Yb_{\{h\}}~\vert~ \Yb_{V^\zeta\backslash\{k,h\}},
\end{align}
meaning that the space $\Yb_{\{k\}}$ and $\Yb_{\{h\}}$ are conditionally independent given $\Yb_{V^\zeta\backslash\{k,h\}}$, 
where $\Phi(e^{i\theta})$ denotes the spectral density matrix of a stochastic vector process $y(t)$,
\begin{align}\nonumber%\label{}
  \Yb_{S} := \Span\{y_{(i)}(t): i\in S,  t\in \Zbb \},
\end{align}
and $S$ is an index set contained in the index set of all entries of $y(t)$. \\  
However, as the complexity and dimensionality of the system increase, 
the underlying vector stochastic process exhibits a low-rank property, 
leading to a rank-deficient spectral density matrix (\cite{lowrankNP24}).
Such processes, referred to as low rank processes,  
have been studied in recent works like \cite{CLP23tac}, \cite{CPL23auto} and \cite{JoeS23}. 
When the vector process underlying a graph or network is low-rank, 
the rank-deficient spectrum violates the well-posedness conditions assumed in existing methods, 
such as those in \cite{Salapaka12tac} and \cite{Materassi21}. 
Moreover, the inverse spectral condition~\eqref{eq:cond_invPhi} can no longer be applied to infer edges between nodes. \\
This paper addresses the lack of a general method for topology recovery of causal low-rank processes through three main contributions. 
First, we introduce a low rank dynamical network (LRDN) model that is computationally efficient
and, crucially, ensures identifiability by guaranteeing a unique graphical representation.
Second, we develop a corresponding topology learning approach based on causal Wiener filtering 
which is theoretically proven to be consistent.
Third, we establish a fundamental one-one correspondence between conditional Granger causality 
and the entries of the causal Wiener filter (or the inverse spectral factor, as shown in \eqref{eq:Sij=0_orth}\eqref{eq:W-1orth}), 
thereby resolving the failure of the conventional inverse spectral condition in \eqref{eq:cond_invPhi}.
Notably, our method imposes no requirement for strictly causal edges in every chain—a key limitation of existing full-rank methods—thus generalizing consistent topology learning to a broader network class.\\
Section~\ref{sec:model} introduces preliminaries and the LRDN model.
Section~\ref{sec:WienerF} derives the causal Wiener filter.
In Section~\ref{sec:approach}, the relation between conditional Granger causality and Wiener filter is discussed, 
and then the topology estimation approach is proposed through 
establishing a sufficient and necessary condition for the existence of edges.
An example and the conclusions are given in Sections~\ref{sec:example} and~\ref{sec:con}.\\
\section{Low rank graphical model}\label{sec:model}
Consider an $(m+l)$-dimensional low rank process,
\begin{align}\nonumber %\label{}
  y(t)= [y_{(1)}(t), y_{(2)}(t), \cdots, y_{(m+l)}(t) ]',
\end{align}
with spectral density matrix $\Phi(z)$ satisfying $\rank(\Phi)=l$. %and $\Phi(\infty)$ constant. 感觉这个没什么必要
Since $\Phi(z)$ is not invertible, 
the conditional dependencies among nodes cannot be directly recovered 
using the conventional inverse spectral condition \eqref{eq:cond_invPhi}.
This paper thus addresses the problem of constructing a graphical structure for low-rank processes, 
where edges reflect inter-nodal relationships and the topology is identifiable.\\
We restrict our attention to causal processes and causal interdependencies, 
which underpin physically realizable identification and control schemes.
We further assume no exogenous inputs, 
so that the graph must be identified solely from non-invasive observations of its internal dynamics.
In the following, we review preliminaries on linear dynamical networks and low rank processes, 
and introduce the formal definition of a low rank dynamical network. 
\subsection{Linear Causal Dynamical Network}
We recall the following foundational definition.
\begin{defn}[Directed and undirected graphs]{\sloppy
A directed (undirected) graph is a pair $(V, E)$, where $V$ is the index set of nodes, and $E$ is a set of edges, 
defined as ordered (unordered) pairs of elements in $V$.}
\end{defn}
%To explore the interrelations among entries of $y(t)$, we give the following definitions of linear causal dynamical networks and graph based on the linear dynamic influence model defined in \cite{Materassi21}.
Denote by $[M]_{ij}$ the $(i,j)$-th entry of a matrix.
Building on the framework in \cite{Materassi21}, we now define a linear causal dynamical network without exogenous inputs and its graph.
\begin{defn}[Linear causal dynamical network model] {\sloppy
A linear causal dynamical network (LCDN) of a process $y(t)$ is defined as a pair $(G(z), w(t))$ where
\begin{itemize}
    \item $G(z)$ is an $(m+l)\times (m+l)$ causal rational matrix,  %where $[G(z)]_{ij}$ denotes its $(i,j)$-th entry, 
    and $[G(\infty)]_{ii}=0$ for $i=1,\cdots, m+l$;
  \item $w(t)=[w_{(1)}(t), \cdots, w_{(m+l)}(t)]'$ is a vector of $m+l$ independent scalar white noises %, %and independent of $y(t)$, 
  with a diagonal spectrum $\Sigma \succeq 0$.
\end{itemize}
The output processes $\{y_{(i)}(t)\}_{i=1}^{m+l}$ of a LCDN are given by
\begin{align}\label{eq:y(i)=w(j)+sumGijy(j)}
y_{(i)}(t)=w_{(i)}(t) + \sum_{j=1}^{m+l} [G(z)]_{ij}y_{(j)}(t), 
\end{align}
or equivalently in vector form
\begin{align}%\label{}
  y(t)= w(t) + G(z)y(t).
\end{align}}
\end{defn} %这里直接定义成causal model
\begin{defn}[Graph associated with a LCDN]{\sloppy
Let $(G(z), w(t))$ be a LCDN with output processes $\{y_{(j)}(t)\}_{j=1}^{m+l}$. Let $V:=\{1,\cdots,m+l\}$ be the node index set, 
and let $E \subseteq V\times V$  such that 
$$(i,j)\in E ~\Leftrightarrow~ [G(z)]_{ij} \neq 0,$$
meaning the edge from $y_{(j)}$ to $y_{(i)}$ exists. 
Then the directed graph $(V,E)$ is the graphical representation of the LCDN.
}\end{defn}
This work extends the traditional linear dynamical influence model in \cite{Materassi21}
by allowing self-loops in the associated graph.
We impose strict causality on the diagonal entries of $G(z)$ by defining $[G(\infty)]_{ii} = 0$,
thereby ensuring that the expansion of the right-hand side (RHS) in \eqref{eq:y(i)=w(j)+sumGijy(j)} contains no term for $y_{(i)}(t)$ at time $t$. \\
%To ensure the identifiability of the resulting LCDN, we impose strict causality on the diagonal entries of $G(z)$ by defining $[G(\infty)]_{ii} = 0$.
%Consequently, the expansion of the right-hand side (RHS) in \eqref{eq:y(i)=w(j)+sumGijy(j)} contains no term for $y_{(i)}(t)$ at time $t$.\\
%
%
While existing methods like \cite{Salapaka12tac}, \cite{Materassi21} and \cite{Lin13}, 
require $\Phi(z)$ to be full-rank for consistent topology identification, 
we relax this constraint to include the more general case of low rank processes and allow self-loops.
Within this extended framework, we investigate the topology learning of low rank processes.
\subsection{Special Feedback Model}\label{subsec:specialFB}
Naturally, suppose $y(t)$ and $\Phi(z)$ can be partitioned as 
\begin{align}\label{eq:partition}
  y(t)=\begin{bmatrix}
         y_{\rm m}(t) \\
         y_{\rm l}(t)
       \end{bmatrix}, ~\Phi(z)=\begin{bmatrix}
                                 \Phi_\mm(z) & \Phi_{\lm\mm}^*(z) \\
                                 \Phi_{\lm\mm}(z) & \Phi_\lm(z)
                               \end{bmatrix},
\end{align}
where $y_\mm(t) \in \Rbb^{m}$, $y_\lm(t)\in \Rbb^{l}$, 
and $\rank(\Phi_\lm(z))=l$ (i.e., $y_\lm$ is a full rank process).
This can be easily achieved by reordering the components of the original process $y^{o}(t)$.\\
Based on the rank-deficient property, from our previous works such as \cite{CLP23tac} and \cite{CPL23auto}, 
a deterministic relation always exists between $\Phi_{{\lm\mm}}(z)$ and $\Phi_{{\lm}}(z)$. 
And the relationship between $y_{\rm m}(t)$ and $y_{\rm l}(t)$ can be described by the following special feedback model.
\begin{defn}[Special feedback model]{\sloppy
For a low rank vector process $y(t)$ as in \eqref{eq:partition} with $\rank(\Phi_\lm(z))=\rank(\Phi(z))=l$,
there is a special feedback model,
\begin{subequations}\label{eq:specialFB}
\begin{align}\label{eq:ym=Hyl}
  y_\mm(t) &= H(z)y_\lm(t),\\  \label{eq:yl=Fym}
  y_\lm(t) &= F(z)y_\mm(t) + v_\lm(t),
\end{align}
\end{subequations}
where $v_\lm(t)$ is an error process. %可能是colored的
The deterministic relation 
\begin{align}\label{eq:H=}
H(z)= \Phi_{{\lm\mm}}(z)^*\Phi_{{\lm}}(z)^{-1}, 
\end{align}
yields a unique $m\times l$-dimensional casual function determined from $\Phi(z)$, 
and $F(z)$ is a (strictly) causal function.
}\end{defn}
The non-unique forward-loop function $F(z)$ may be determined from such as a minimal realization of $y(t)$ (\cite{CLP23tac}),
or a one-step Wiener predictor (\cite{CPL23auto}). 
\subsection{Low Rank Dynamical Network}\label{subsec:LRgraphs}
Though $y(t)$ is low-rank with a singular spectrum, 
its subprocess $y_\lm(t)$ has a nonsingular spectrum.
Recalling the causal deterministic relation from $y_\lm$ to $y_\mm$, 
a specific LCDN can be constructed for $y(t)$ via direct calculations.
\begin{prop}\label{prop:lowrankLCDN}{\sloppy
The low rank process $y(t)$ admits a LCDN representation $(G(z), w(t))$ with
\begin{align}%\label{}
  G(z) = \begin{bmatrix}
           0 & G_{\mm\lm}(z) \\
           0 & G_\lm(z)
         \end{bmatrix},~ w(t)=\begin{bmatrix}
                                0 \\
                                w_\lm(t) 
                              \end{bmatrix},
\end{align}
where $G_{\mm\lm}(z), G_\lm(z)$ are causal rational, and $w_\lm(t)$ is $l$-dimensional with a full-rank spectrum.
}\end{prop}
From Proposition~\ref{prop:lowrankLCDN}, the edge set $E$ of the corresponding graph can be shrunk to a subset of $V \times \{ m+1, \cdots, m+l \}$, 
because $[G(z)]_{ij}\equiv 0$ for $j=1, \cdots m$.
To simplify the representation and the subsequent topology estimation, 
we introduce the following model.
\begin{defn}[Low rank dynamical network model]\label{def:LDRN}{\sloppy
A low rank dynamical network (LRDN) for process $y(t)$ under partition \eqref{eq:partition} with $\rank(\Phi_\lm(z))=\rank(\Phi(z))=l$,
is a triple $(G_{\mm\lm}(z), G_\lm(z), w_\lm(t))$, where
\begin{itemize}
    \item $G_{\mm\lm}(z)$ is an $m\times l$ causal rational matrix; %, determined by \eqref{eq:H=}.
    \item $G_\lm(z)$ is an $l \times l$ causal rational matrix, where %$[G_\lm(z)]_{ij}$ denotes its $(i,j)$-th entry, and 
    $[G_\lm(\infty)]_{ii}=0$ for $i=1,\cdots,l$;
    \item $w_\lm(t)=[w_{\lm(1)}, \cdots, w_{\lm(l)}]'$ is a vector of $l$ independent scalar white noises with a diagonal spectrum $\Sigma_\lm \succ 0$.
\end{itemize}
The output processes are given by %\eqref{eq:ym=Hyl}\eqref{eq:yl=Glyl}, i.e.,
\begin{subequations}\label{eq:LRDN}
\begin{align}\label{eq:ym=Gmlyl}
  y_\mm(t) &= G_{\mm\lm}(z)y_\lm(t), \\ \label{eq:yl=w+Glyl}
  y_\lm(t) &= w_\lm(t) + G_\lm(z)y_\lm(t). 
\end{align}
\end{subequations}
}\end{defn}
\begin{defn}[Graph associated with a LRDN]\label{def:graphLDRN}{\sloppy
Let $(G_{\mm\lm}(z), G_\lm(z), w_\lm(t))$ be a LRDN of process $y(t)$ with outputs in \eqref{eq:LRDN}. 
Let $V:=\{1,\cdots,m+l\}$ be the node index set and $V_{\lm}:=\{ m+1, \cdots, m+l \}$.
Let $E = E_\mm \cup E_\lm$, $E_\mm \subset (V\setminus V_\lm) \times V_\lm$, $E_\lm \subset V_\lm \times V_\lm $, 
such that for $j\in V_\lm$, 
\begin{subequations}\label{eq:edges}
\begin{align}
   (i,j)\in E_\mm ~&\Leftrightarrow~ [G_{\mm\lm}(z)]_{i,j-m} \neq 0, {\rm ~for~} i \in (V\setminus V_\lm), \\
   (i,j)\in E_\lm ~&\Leftrightarrow~ [G_\lm(z)]_{i-m,j-m} \neq 0, {\rm ~for~} i\in V_\lm.
\end{align}
\end{subequations}
Then the directed graph $(V,E)$ is the graphical representation of the LRDN.
}\end{defn}
Definitions~\ref{def:LDRN} and~\ref{def:graphLDRN} establish a 
one-one correspondence between an LRDN and a low-rank process under a fixed partition (the proof of this identifiability is omitted for space). 
The LRDN is well-posed if $I-G_\lm(z)$ is invertible. %这里也比之前的论文放宽了条件
The associated LRDN graph contains at most $(m+l) \times l$ edges, compared to $(m+l)^2$ for a full LCDN graph. 
This sparsity leads to significant computational advantages. 
In the following sections, we develop a Wiener filter-based method for estimating the unique graph of a LRDN.
\section{Causal Wiener filter}\label{sec:WienerF} %给出维纳滤波表达式
While the causal deterministic relation from $y_\lm$ to $y_{\mm}$ in \eqref{eq:ym=Hyl}
determines the edges in $E_{\mathrm{m}}$, 
this section, focuses on the internal topology within $y_\lm$.
We derive a Wiener filter for each entry of $y_\lm(t)$ based on the other entries and its own strict past,
and establish the uniqueness of this filter regardless of the specific spectral factor.\\
%
%\subsection{Wiener Filter in $y_\lm$}
%
Suppose the full-rank process $y_\lm(t)$ admits a minimum-phase spectral factorization
%\begin{subequations}\label{eq:specfact}
\begin{align}\label{eq:Phil=WW*}
 \Phi_\lm(z)= W(z)\Lambda W(z)^*,
\end{align}
where $W(z)$ is $l$-dimensional and square, with the form
\begin{align}\label{eq:W=sumWkz-k}
    W(z)=\sum_{k=0}^{n_W} W_k z^{-k},
\end{align}
%\end{subequations}
for some degree $n_W$, $W_k$ are constant matrices, $W_0$ is invertible.
Suppose the corresponding $l$-dimensional innovation process is $e(t)$ with spectrum $\Lambda$, we have
\begin{align}\label{eq:yl=We}
 y_\lm(t)=W(z)e(t).
\end{align}
Under the above settings, 
$W(z)$ is unique up to right multiplication by a constant unitary matrix $U$ satisfying $\Lambda U=U\Lambda$.\\
Denote the $i$-th entry of $y_{\lm}(t)$ by $y_{\lm(i)}(t) = y_{(i+m)}(t)$. 
For a subset $\Ical = \{i_1, i_2, \cdots, i_r \} \subseteq \{1, \cdots,l \}$ satisfying $i_1< \cdots <i_r$,
denote by $B_\Ical=[b_{i_1}, b_{i_2}, \cdots, b_{i_r} ]'$, 
with $b_{k}$ the $k$-th elementary (column) vector with $1$ in the $k$-th entry and $0$ elsewhere.
Denote by
\begin{align}\nonumber
 y_{\lm \Ical}(t) := B_\Ical y_\lm(t), \quad
 W_{\Ical}(z) := B_\Ical W(z),
\end{align}
the entries of $y_\lm(t)$ indexed by $\Ical$, and the rows of $W(z)$ indexed by $\Ical$.
Let $\overline{\{\Ical\}}:=\{1, \cdots,l \} \setminus \{\Ical\}$,
and let $\{ \cdot \}_\Ccal$ denote the orthogonal projection operator yielding the causal part of a function.
Then we have the following results.
\begin{lem}\label{lem:hatyl(i)}
Suppose the full rank process $y_\lm(t)$ has a spectral factorization \eqref{eq:Phil=WW*}. %,
%with $\Lambda$ diagonal and fixed.
For process $y_{\lm(i)}(t)$, define subspace 
\begin{equation}\label{eq:Yl(i)-}%\nonumber
\begin{split}
\Yb_{\lm(i)}^-(t):= \Span \{   y_{\lm(i)} (k-1),~y_{\lm(j)}(k): 
  j\in\overline{\{i\}}, k\leq t \},
\end{split}
\end{equation}
hereafter $\Yb_{\lm(i)}^-$, meaning the space generated by the strict past of $y_{\lm(i)}$ 
and the past of other entries in $y_\lm$.
Consider the problem 
\begin{align}\nonumber %\label{}
    \min_{q(t)\in \Yb_{\lm(i)}^-} \Vert  q(t)- {y}_{\lm(i)}(t) \Vert^2, 
\end{align}
for a Wiener filter for ${y}_{\lm(i)}(t)$ on space $\Yb_{\lm(i)}^-$.
The solution exists, is unique regardless of the specific $W(z)$, and has the form 
\begin{subequations}\label{eq:WienerF_i}
\begin{align}\label{eq:hatyl(i)}
\hat{y}_{\lm(i)}(t) = (W_{\{i\}}(z)- [W_0]_{ii}b_i')W(z)^{-1}M_i(z)^* x_i^-(t),
\end{align}
where 
\begin{align}\label{eq:Mi}
    M_i(z) := I+(z^{-1}-1)b_ib_i', 
\end{align}
\begin{align}\label{eq:xi+}
\begin{split}
   x_i^-(t) := &\begin{bmatrix}
                    B_{\overline{\{i\}}} \\
                    b_{i} 
                  \end{bmatrix}'\begin{bmatrix}
                    y_{\lm \overline{\{i\}}} (t) \\
                    y_{\lm (i)} (t-1)
                  \end{bmatrix}  \\
   = & \left[ y_{\lm(1)}(t), \cdots, y_{\lm (i)} (t-1) \cdots, y_{\lm(l)}(t) \right]'.
\end{split}
\end{align}
\end{subequations}
\end{lem}

\begin{pf}
From \eqref{eq:Mi}, $M_i(z)$ is a diagonal causal function matrix with entry $[M_i(z)]_{ii}=z^{-1}$ and other diagonal entries equal to 1.
It follows directly that 
$$
M_i(z)^*=I+(z-1)b_ib_i',
$$
with $[M_i(z)]_{ii}=z$, and $M_i(z)M_i(z)^*=I$, i.e., $M_i(z)$ is unitary.
Then $x_i^-(t)$ in \eqref{eq:xi+} is given by 
\begin{align}\label{eq:xi+=Miyl}
   x_i^-(t) = M_i(z) y_\lm(t).
\end{align}
Denote by 
$$
\varepsilon_i^-(t) := M_i(z)e(t).
$$
Recall from \eqref{eq:yl=We} that an innovation process of $x_i^-(t)$ is given by $P\varepsilon_i^-(t)$
with an arbitrary $l\times l$ constant unitary matrix $P$ satisfying $P^*P=I$. 
Consider the general solution $W(z)U$ to \eqref{eq:Phil=WW*} satisfying $\Lambda U= U\Lambda$ and $UU^*=I$, 
such that
\begin{align}\nonumber %\label{}
  x_i^-(t) = M_i(z) W(z) UU^* M_i(z)^* P^* P\varepsilon_i^-(t).
\end{align}
Since
$$ 
y_{\lm(i)}=W_{\{i\}}(z) UU^* M_i(z)^* P^* P\varepsilon_i^-(t), 
$$ 
the Wiener filter for $y_{\lm(i)}(t)$ on $\Yb_{\lm(i)}^-$ is given by
\begin{align}\nonumber%\label{}
 \E \{ y_\lm(t) \vert \Yb_{\lm(i)}^- \} = \{W_{\{i\}}M_i^*P^* \}_\Ccal \left(M_i W M_i^* P^*\right)^{-1} x_i^-(t),
\end{align}
which keeps the same for different spectral factors satisfying \eqref{eq:Phil=WW*}.
We now prove this Wiener filter is unique independent of the specific choice of $P$.
%and has the form  \eqref{eq:hatyl(i)}. 
Since $P^*$ is constant, the entries of 
$W_{\{i\}}M_i^*P^*$ corresponds to $l$ linear combinations of the entries in following row-vector,
\begin{align}\nonumber 
W_{\{i\}}M_i^* = \left[ [W]_{i1}, \cdots, z[W]_{ii}, \cdots, [W]_{il} \right],
\end{align}
where only the $i$-th entry might be not causal because $W(z)$ is causal.
Thus the causal projection and linear combination operations can commute, i.e.,
$\{W_{\{i\}}M_i^*P^* \}_\Ccal = \{W_{\{i\}}M_i^*\}_\Ccal P^*$, and 
\begin{align}\nonumber%\label{}
\begin{split}
 \E \{ y_\lm(t) \vert \Yb_{\lm(i)}^- \} &= \{W_{\{i\}}M_i^* \}_\Ccal P^* \left(M_i W M_i^* P^*\right)^{-1} x_i^-(t) \\
 &= \{W_{\{i\}}M_i^* \}_\Ccal M_iW^{-1}M_i^* x_i^-(t),
\end{split}
\end{align}
verifying the uniqueness.\\
From \eqref{eq:W=sumWkz-k}, 
\begin{align}\nonumber \begin{split}
 \{z[W(z)]_{ii}\}_\Ccal %&= \{ z\sum_{k=0}^{n_W} [W_k]_{ii}z^{-k} \}_\Ccal \\
 &=\{ z[W_0]_{ii} + \sum_{k=1}^{n_W}[W_k]_{ii}z^{-k+1}  \}_\Ccal \\
 &= z[W(z)]_{ii} - z[W_0]_{ii}.
\end{split}\end{align}
Then we have
\begin{align}\nonumber \begin{split}
  \{W_{\{i\}}M_i^* \}_\Ccal M_i &= (W_{\{i\}}M_i^* - z[W_0]_{ii}b_i' )M_i \\
  &=W_{\{i\}}(z) - [W_0]_{ii}zb_i'M_i.
\end{split}\end{align}
Combined with \eqref{eq:Mi}, the unique Wiener filter is simplified to \eqref{eq:hatyl(i)}.
\hfill $\square$
\end{pf}
From \eqref{eq:xi+=Miyl}, we have $M_i(z)^*x_i^-(t) = y_l(t)$. 
Then from \eqref{eq:WienerF_i} we have
\begin{align}\nonumber
\begin{split}
  \begin{bmatrix}
    \hat{y}_{\lm(1)}(t) \\
    \vdots \\
    \hat{y}_{\lm(l)}(t) 
  \end{bmatrix}&= \begin{bmatrix}
                    (W_{\{1\}}(z)- [W_0]_{11}b_1')W(z)^{-1}y_\lm(t) \\
                    \vdots \\
                    (W_{\{l\}}(z)- [W_0]_{ll}b_2')W(z)^{-1}y_\lm(t)
                  \end{bmatrix}, %\\ 
\end{split}\end{align}
which can be concluded by the following theorem (with the proof omitted for the space).
\begin{thm}\label{thm:WienerF}
Suppose the full rank process $y_\lm(t)$ has a spectral factorization \eqref{eq:Phil=WW*}. 
Denote by 
\begin{align}\label{eq:hatyl_def}
    \hat{y}_\lm(t)= \left[ \hat{y}_{\lm(1)}(t), \cdots,  \hat{y}_{\lm(l)}(t)  \right]',
\end{align}
where $\hat{y}_{\lm(i)}(t)$ is the Wiener filter for ${y}_{\lm(i)}(t)$ on space $\Yb_{\lm(i)}^-$ from Lemma~\ref{lem:hatyl(i)}
for $i=1, \cdots, l$.
Then there is a mapping $S(z): \Yb_{V_\lm}^- \rightarrow \Yb_{V_\lm}^- $
uniquely determined from the Wiener filters $\hat{y}_{\lm(i)}(t)$ for $i=1, \cdots, l$, such that
\begin{subequations}\label{eq:WienerF}
\begin{align}
 \hat{y}_{\lm}(t) = S(z)y_{\lm}(t),
\end{align}
and $S(z)$ has the form
\begin{align}\label{eq:S=I-DWinv}
S(z) = I-DW(z)^{-1},
\end{align}
where $D$ is a diagonal matrix satisfying
\begin{align}
    [D]_{ii} = [W_0]_{ii}, \quad{\rm for~} i=1, \cdots, l,
\end{align}
with $W_0$ given in \eqref{eq:W=sumWkz-k}.
\end{subequations}
Moreover, $S(z)$ is unique regardless of the specific $W(z)$,
and $[S(\infty)]_{ii}=0$ for $i=1, \cdots, l$.
\end{thm}
Consequently, 
the causal Wiener filter for each node based on its strict past and the past of other nodes, 
is uniquely determined. 
The next section establishes the equivalence between this filter and conditional Granger causality,
thereby enabling the reconstruction of the directed network topology that encodes these causal relationships.
\section{Topology learning by Wiener Filter}\label{sec:approach} %证明拓扑对每个LCDN都存在且唯一，给出重构伪代码
Building on the unique Wiener filter derived above, 
we now formalize its connection to Granger causality and topology learning.
\subsection{Conditional Granger Causality}
We first recall the geometric definition of Granger non-causality (\cite{CLP23tac}):
there is no causality in the sense of Granger from process $\beta$ to $\alpha$ if and only if
\begin{align}%\label{}
  \Ab^+  \perp \Bb^- \vert \Ab^{--},
\end{align}
where $\Ab^{--}$, $\Ab^+$ are subspaces generated by the strict past and the future (including present) of $\alpha(t)$, respectively, 
and $\Bb^-$ is the subspace generated by the past of $\beta(t)$.
Similarly, strict non-causality is equivalent to $\Ab^+  \perp \Bb^{--} \vert \Ab^{--}$, 
where $\Bb^{--}$ denotes the strictly past subspace of $\beta(t)$.
This geometric formulation leads us to the following definition of the conditional case.
\begin{defn}
For stochastic processes $\alpha(t), \beta(t), \gamma(t)$, we say 
there is no conditional Granger causality from $\beta$ to $\alpha$ given $\gamma$, 
if and only if 
\begin{align}\label{eq:cond2_Granger}
    \Ab^+  \perp \Bb^- \vert \Ab^{--} \vee \Cb^{-}.
\end{align}
or equivalently, for any $\lambda \in \Ab^{+}$,
\begin{align}\label{eq:cond1_Granger}
\E \{ \lambda \vert \Bb^{-} \vee \Ab^{--} \vee \Cb^{-} \} = \E \{ \lambda \vert \Ab^{--} \vee \Cb^{-} \},
\end{align}
where $\Cb^-$ is subspace generated by the past of $\gamma(t)$.
\end{defn}
The equivalence between \eqref{eq:cond2_Granger} and \eqref{eq:cond1_Granger} follows from \cite{LPbook} (Proposition 2.4.2).
The following theorem establishes a necessary and sufficient condition 
in terms of the zero entries of the Wiener filter, for conditional non-causality between different nodes.
\begin{thm}\label{thm:SijGranger}
Define the subspaces 
\begin{subequations}\label{eq:subspaces}
\begin{align}
&\Yb_{\lm \Ical}^-(t) := \Span \{ y_{\lm(h)}(k):  h\in \Ical, k\leq t \},\\ 
&\Yb_{\lm\{i\}}^{--}(t):= \Span \{ y_{\lm(i)}(k), k<t \}, \\ 
& \Yb_{\lm\{i\}}^+(t) := \Span \{ y_{\lm(i)}(k), k\geq t \},
\end{align} 
\end{subequations}
hereafter $\Yb_{\lm\Ical}^-, \Yb_{\lm\{i\}}^{--}, \Yb_{\lm\{i\}}^+$.
Let $S(z)$ be the Wiener filter matrix from Theorem~\ref{thm:WienerF}.
Then, for any $i\neq j$,
\begin{align}\label{eq:Sij=0_orth}
[S(z)]_{ij} = 0  
 \Leftrightarrow  \Yb_{\lm\{i\}}^+ \perp \Yb_{\lm\{j\}}^- ~\vert~ (\Yb_{\lm\{i\}}^{--} \vee \Yb_{\lm \overline{\{i,j\}}}^- ),
\end{align}
that is, $[S(z)]_{ij}=0$ if and only if there is no conditional Granger causality from $y_{\lm(j)}(t)$ to $y_{\lm(i)}(t)$ 
given the past of all other entries in $y_\lm(t)$. 
\end{thm}
\begin{pf}\sloppy{
We first prove that for any $i\neq j$,
\begin{align}\label{eq:Sij=0_orth1}
\begin{split}
 [S(z)]_{ij} = 0  
 \Leftrightarrow  y_{\lm(i)}(t) \perp \Yb_{\lm\{j\}}^- ~\vert~ (\Yb_{\lm\{i\}}^{--} \vee \Yb_{\lm \overline{\{i,j\}}}^- ),
\end{split}
\end{align} 
and then extends this to \eqref{eq:Sij=0_orth}.\\
Let $\Zb(t):= \Yb_{\lm\{i\}}^{--}(t) \vee \Yb_{\lm \overline{\{i,j\}}}^-(t)$.
The RHS of \eqref{eq:Sij=0_orth1} is equivalent to 
\begin{align}\label{eq:pf_orthcond}
\begin{split}
  \E\{ y_{\lm(i)}(t) \vert \Yb_{\lm(i)}^-(t) \} & = \E\{ y_{\lm(i)}(t) \vert \Zb(t) \},
\end{split}
\end{align}
with $\Yb_{\lm(i)}^-$ defined in \eqref{eq:Yl(i)-}.
%Denote $S_{\{i\}}$ the $i$-th row of $S(z)$, the LFS of \eqref{eq:pf_orthcond} is
From Theorem~\ref{thm:WienerF},
the left-hand side (LFS) of \eqref{eq:pf_orthcond} is
\begin{align}\nonumber
\E\{ y_{\lm(i)}(t) \vert \Yb_{\lm(i)}^-(t)  \}= \hat{y}_{\lm(i)}(t) = \sum_{h=1}^{l} [S(z)]_{ih} y_{\lm(h)}(t).
\end{align}
Denote by $e_{(i)}(t)$ the $i$-the entry of $e(t)$. Then from \eqref{eq:WienerF},
$$
 {y}_{\lm(i)}(t) - \E\{ {y}_{\lm(i)}(t) \vert \Yb_{\lm(i)}^-(t)  \} = [D]_{ii}e_{(i)}(t).
$$
{\bf Sufficiency} of \eqref{eq:Sij=0_orth1}: Suppose $[S(z)]_{ij}=0$.
Since $\Zb(t) \subseteq \Yb_{\lm(i)}^-(t) $, 
we have 
\begin{align}\label{eq:pf_ineq}
\begin{split}
&\Vert {y}_{\lm(i)}(t)- \E\{ {y}_{\lm(i)}(t) \vert \Zb(t) \} \Vert^2  \\
\geq & \Vert {y}_{\lm(i)}(t)- \E\{ {y}_{\lm(i)}(t) \vert \Yb_{\lm(i)}^-(t) \}\Vert^2
= \Vert [D]_{ii}e_{(i)}(t) \Vert ^2.
\end{split}
\end{align}
Moreover, $\hat{y}_{\lm(i)}(t) \in  \Zb(t) $ from $[S(z)]_{ij}=0$ and $y_{\lm(i)}(t)-\hat{y}_{\lm(i)}(t) = [D]_{ii}e_{(i)}(t) \perp \Zb(t)$. 
Combining this with $\eqref{eq:pf_ineq}$, it follows that $\hat{y}_{\lm(i)}$ is also the Wiener filter of ${y}_{\lm(i)}$ on $\Zb(t)$, 
and thus \eqref{eq:pf_orthcond} holds. \\
{\bf Necessity} of \eqref{eq:Sij=0_orth1}: Suppose \eqref{eq:pf_orthcond} holds. 
Then,
$$
 \sum_{h=1}^{l} [S(z)]_{ih} y_{\lm(h)}(t)  \in  \Zb(t).
$$
Since $\sum_{h\neq j} [S(z)]_{ih} y_{\lm(h)}(t)\in  \Zb(t)$, 
it follows that $[S(z)]_{ij} y_{\lm(j)}(t) \in \Zb(t)$. 
This implies
$$
 [S(z)]_{ij} y_{\lm(j)} (t) = T(z) y_{\lm\overline{\{i,j\}}}(t),
$$
where $T(z)$ is a causal polynomial row vector. 
If $T(z)\neq 0$ and hence $[S(z)]_{ij}\neq 0$, 
the Wiener filter $\hat{y}_{\lm(i)}(t)$ would be non-unique,  
contradicting the uniqueness in Lemma~\ref{lem:hatyl(i)}.
Hence, $[S(z)]_{ij}= 0$ is necessary.\\
{\bf Extension} to \eqref{eq:Sij=0_orth}: the RHS of \eqref{eq:Sij=0_orth1} is equivalent to
$$
y_{\lm(i)}(t+\tau) \perp \Yb_{\lm\{j\}}^-(t+\tau) ~\vert~ \Zb(t+\tau),
$$
for any $\tau\geq0$. Since $\Zb(t) \subseteq  \Zb(t+\tau)$, and $\Yb_{\lm\{j\}}^- (t) \subseteq \Yb_{\lm\{j\}}^-(t+\tau)$
%$$
%  \Zb(t) \subseteq  \Zb(t+\tau),~ \Yb_{\lm\{j\}}^- (t) \subseteq \Yb_{\lm\{j\}}^-(t+\tau)
%$$
for $\tau \geq 0$, it follows that
\begin{align}\label{eq:pf_last}
y_{\lm(i)}(t+\tau) \perp \Yb_{\lm\{j\}}^-(t) ~\vert~ \Zb(t).
\end{align}
Now, for any $\lambda = \bar{T}(z)y_{\lm(i)}(t)\in \Yb_{\lm\{i\}}^+$ with $\bar{T}(z) = \sum_{k=0}^{n_T} \bar{T}_kz^{k}$ anticausal, 
we have
$$
\lambda = \sum_{k=0}^{n_T} \bar{T}_k y_{\lm(i)}(t+k),
$$
where each term $\bar{T}_k y_{\lm(i)}(t+k)$ is conditionally orthogonal to subspace $\Yb_{\lm\{j\}}^-(t)$ given $\Zb(t)$ from \eqref{eq:pf_last}.
Therefore,
$$
 \lambda \perp \Yb_{\lm\{j\}}^-(t) ~\vert~ \Zb(t),  ~\forall \lambda \in \Yb_{\lm(i)}^+,
$$
which establishes \eqref{eq:Sij=0_orth}.
\hfill $\square$
}\end{pf}
From Theorem~\ref{thm:SijGranger} and Equation \eqref{eq:S=I-DWinv}, the following result is ready
concerning conditional Granger non-causality between nodes in a full-rank vector process, 
which provides a criterion for causal inference without explicit reference to the graph topology.
\begin{cor}\label{cor:W-1}
For a full-rank process $y_\lm(t)$ with a minimum-phase spectral factor $W(z)$ in \eqref{eq:Phil=WW*},
and for $i\neq j$, we have
\begin{align}\label{eq:W-1orth}
\begin{split}
 [W(z)^{-1}]_{ij} = 0 \Leftrightarrow  \Yb_{\lm\{i\}}^+ \perp \Yb_{\lm\{j\}}^- ~\vert~ (\Yb_{\lm\{i\}}^{--} \vee \Yb_{\lm \overline{\{i,j\}}}^- ), 
\end{split}
\end{align}
$\Leftrightarrow$ the absence of~conditional~Granger~causality~from~ $y_{\lm(j)}$ to $y_{\lm(i)}$ given the rest entries of $y_\lm(t)$.
\end{cor}
In contrast to the symmetric relation of conditional independence characterized by \eqref{eq:cond_invPhi}, 
conditional Granger non-causality is directional and typically asymmetric.
This directional property enables the direct learning of oriented edges. 
It bypasses the complexity inherent in inferring directed connections from symmetric measures, 
which often requires excluding indirectly connected `kin nodes'.
We thus proceed to reconstruct the directed causal graph associated with a LRDN, 
where the presence of an edge $(i,j)$ is equivalent to the existence of conditional Granger causality from $y_{\mathrm{l}(j)}$ to $y_{\mathrm{l}(i)}$.
\subsection{LRDN Topology Learning}   
As established in Subsection~\ref{subsec:specialFB}, matrix 
$G_{\mm\lm}(z)$ in a LRDN \eqref{eq:LRDN} can be uniquely determined by 
the causal deterministic relation $H(z)$ in the special feedback model \eqref{eq:specialFB},
which is itself a Wiener filter. 
We now demonstrate that the entire topology of an LRDN can be exactly reconstructed using Wiener filters between nodes. 
The associated graph admits both loops and strictly causal self-loops, 
offering greater generality than existing approaches. 
Moreover, the presence of a directed edge precisely corresponds to conditional Granger causality between the nodes.
\begin{thm}\label{thm:GrangerLDRN}
For a LRDN $(G_{\mm\lm}(z), G_\lm(z), w_\lm(t))$, 
suppose the $(m+l)$-dimensional output process $y(t)$ with its spectrum are given in \eqref{eq:partition}, 
satisfying $\rank(\Phi(z)) = \rank(\Phi_\lm(z))=l$. 
Define subspaces $\Yb_\Ical^-$, $\Yb_{\{i\}}^{--}$, $\Yb_{\{i\}}^+$ as in \eqref{eq:subspaces} 
by substituting $y_\lm(t)$ by $y(t)$.\\
Then, the edge set $E$ of the graph $(V, E)$ associated with the LRDN can be consistently 
reconstructed by the following sufficient and necessary condition.
For $j \in V_\lm$,
\begin{align}\label{eq:topo<=>}
\begin{split}
  (i,j) \in E ~\Leftrightarrow~ 
  \left\{ \begin{array}{ll}
                                               {[H(z)]_{i,j-m}\neq0,} & {i\in (V\setminus V_\lm),} \\
                                               {[S(z)]_{i-m,j-m}\neq0,} & {i\in V_\lm,} %\\
                                               %{[S(z)-S(\infty)]_{i-m,i-m}\neq 0}, & {i\in V_\lm,}
                                             \end{array}\right.
\end{split}
\end{align}
where $V=\{1, \cdots, m+l\}$, $V_\lm =\{m+1, \cdots, m+l \}$,
$H(z)y_\lm(t)$ is the causal Wiener filter for $y_\mm(t)$ given $y_\lm(t)$,
the $i$-th row of $S(z)y_\lm(t)$, denoted by $S_{\{i\}}(z)y_\lm(t)$, 
is the causal Wiener filter of $y_{\lm(i)}(t)$ given its own strict past and the past of all other entries in $y_\lm(t)$. \\
Moreover, for a LRDN, there is no conditional Granger causality from node $y_{(j)}(t)$ to node $ y_{(i)}(t)$ given all other entries in $y(t)$,
if and only if $(i,j) \not\in E$, i.e.,
\begin{align}%\label{}
  (i,j) \not\in E \Leftrightarrow \Yb_{\{i\}}^+ \perp  \Yb_{\{j\}}^- \vert \Yb_{\{i\}}^{--}  \vee \Yb_{V\setminus \{i,j\}}^{-},
\end{align}
for $i\in V$, and $j\in V_\lm$.
\end{thm}
\begin{pf}{\sloppy
The LRDN definition relies on the existence of a deterministic causal relation exists from $y_\lm$ to $y_\mm$. 
As shown in Subsection~\ref{subsec:specialFB} and \eqref{eq:H=},
$H(z)$ is the unique causal Wiener filter from $y_\lm$ to $y_\mm$. 
Therefore, in the LRDN we have
\begin{align}\label{eq:Gml=H}
G_{\mm\lm}(z)=H(z),
\end{align}
which implies that the non-zeros entries of $H(z)$ correspond exactly to the edges in $E_\mm \subset (V\setminus V_\lm) \times V_\lm$. \\
With the same innovation spectrum $\Lambda=\Sigma_\lm\succ 0$,
it follows from \eqref{eq:yl=w+Glyl}\eqref{eq:S=I-DWinv} that 
$$
I-S(z) = D(I-G_\lm(z)).
$$
Consequently,
\begin{align}\nonumber %\label{}
 & [S(z)]_{ij}=[D]_{ii}[G_\lm(z)]_{ij}, {\rm~ when~}i\neq j, \\ \nonumber
 & 1-[S(z)]_{ii} = [D]_{ii}(1-[G_\lm(z)]_{ii}).
\end{align}
Since $[G_\lm(\infty)]_{ii}=[S(\infty)]_{ii}= 0$, 
we have $[G_\lm(z)]_{ii} \neq 0$ $\Leftrightarrow$ $[G_\lm(z)]_{ii}$ is strictly causal nonzero 
$\Leftrightarrow$ $[S(z)]_{ii}$ is strictly causal and non-zero.
Therefore, by \eqref{eq:edges}, the edge condition \eqref{eq:topo<=>} holds.\\
We now prove the necessary and sufficient condition for conditional Granger non-causality, 
considering $i \leq m$ and $i > m$ separately. 
Since both $y_\mm$ and $y_\lm$ are determined by $y_\lm$ in \eqref{eq:LRDN},
the conditional Granger non-causality from $y_{(j)}$ to $y_{(i)}$ for $j\in V_\lm$ is equivalent to 
\begin{subequations}
\begin{align}\label{eq:pf_cond1}
 &\Yb_{i}^+ \perp \Yb_{j}^{-} \vert  \Yb_{\lm\overline{\{j-m\}}}^-, {\rm~for~} i\leq m,\\ \label{eq:pf_cond2}
 &\Yb_{i}^+ \perp \Yb_{j}^{-} \vert \Yb_{\{i\}}^{--} \vee \Yb_{\lm\overline{\{i-m,j-m\}}}^-, {\rm~for~} i> m.
\end{align}
\end{subequations}
For $i>m$, by Theorem~\ref{thm:SijGranger}, \eqref{eq:pf_cond2} holds 
$\Leftrightarrow$ $[S(z)]_{i-m,j-m}=0$ $\Leftrightarrow (i,j) \not\in E_\lm$.
For $i\leq m$, note that $\Yb_{\{j\}}^- \vee \Yb_{\lm\overline{\{j-m\}}}^{-} = \Yb_\lm^-$. 
From \eqref{eq:ym=Gmlyl}\eqref{eq:Gml=H}, 
\begin{align}\nonumber
%\begin{split}
 \E\{ y_{\lm(i)}(t) \vert  \Yb_{\{j\}}^- \vee \Yb_{\lm\overline{\{j-m\}}}^{-}   \} 
 = \sum_{h=1}^{l} [H(z)]_{ih} y_{\lm (h)}. 
%\end{split}
\end{align}
Following a derivation similar to that in the proof of Theorem~\ref{thm:SijGranger},
we conclude that \eqref{eq:pf_cond1} $\Leftrightarrow$ $[H(z)]_{i,j-m}=0$ $\Leftrightarrow (i,j) \not\in E_\mm$.
\hfill $\square$
}\end{pf}
%
%
%From Theorem~\ref{thm:GrangerLDRN}, 
%the algorithm to topology reconstruction of a LRDN with a low rank underlying process \eqref{eq:partition} 
%can be concluded in Algorithm~\ref{alg:topo}.
%
%\begin{algorithm}
%\caption{Topology reconstruction of low rank dynamical network (LRDN)}
%\label{alg:topo}
%\begin{algorithmic}[1]
%\Require {$V$, $V_\lm$, $\{y(k)\}_{k=1}^{N}$}
%\Ensure {Topology}
%\State 3
%\end{algorithmic}
%\end{algorithm}
%
%
The graphical structure of a LRDN estimated in Theorem~\ref{thm:GrangerLDRN},
fully captures the conditional Granger causal relationships between different nodes in the low rank output process. 
This Wiener filter-based formulation offers significant identification advantages when $W_0$ is set to have unit diagonal entries. 
In this framework, $H(z)$ directly represents the deterministic relation, 
while $S(z)$ links to a minimum-phase spectral factor. 
Further elaboration is reserved for the journal version of this work. 
\section{Simulation Example}\label{sec:example}
\begin{figure}%[t]
%\centering
\hspace{-1.2mm}
\subfigure[Unique graph of LRDN]{\hspace{-2mm}
        \includegraphics[scale=0.37]{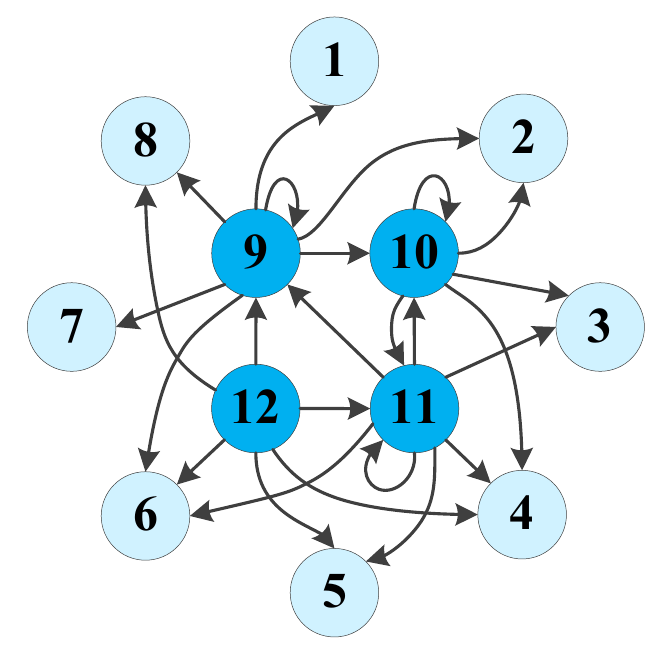}} 
\subfigure[Possible graph of LCDN]{\hspace{0mm}
        \includegraphics[scale=0.37]{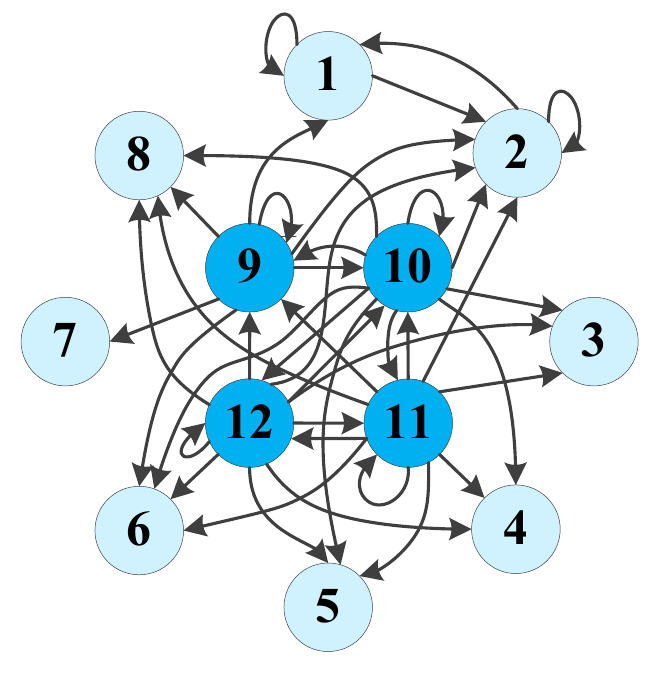}} 
\caption{Graphs associated to different network models of the low rank process in the simulation example. 
Panel (a) also corresponds to the estimated LRDN graph by Theorem~\ref{thm:GrangerLDRN}. }\label{fig:topo}
\end{figure}

We consider a $12$-dimensional process
$y(t)$ innovated by a $4$-dimensional white noise with identify spectrum,
generated as
$$
y(t)=\begin{bmatrix}
       H(z)W(z) \\
       W(z) 
     \end{bmatrix}e(t).
$$
To construct the graph, 
we first define the edge set $E_{\mathrm{l}}$ by specifying the off-diagonal nonzero entries of $W(z)^{-1}$ according to Corollary~\ref{cor:W-1}, 
and set $y_{(12)}(t) = e_{(4)}(t)$ so that node $12$ has no self-loop or incoming edge in $E_{\mathrm{l}}$.  
Next, we form $E_{\mathrm{m}}$ by choosing the nonzero entries of $H(z)$. 
The overall edge set is $E = E_{\mathrm{m}} \cup E_{\mathrm{l}}$. 
%By choosing the non-zeros entries of $H(z)$, we generate $E_\mm$.
Finally, we randomly generate values of the non-zero entries of $W(z)$ and $H(z)$ consistent with $E$, and simulate $y(t)$ for $t=1, \cdots, 200$. \\
The graph of the LRDN generated for this example 
is shown in Fig.~\ref{fig:topo} (a),
where the nodes in $y_\lm$ are marked in darker blue. 
By modeling as a LRDN, there are 25 directed edges among the 12 nodes in its unique graph, 
pointing from $V_\lm$ to $V$. 
This graph represents the target topology to be estimated from the simulated data.\\
In this case, the singular spectrum prevents the construction of an interactive graph (i.e., moral graph)
from conditional dependencies between different nodes. 
Ignoring the low-rank nature of the process during reconstruction leads to non-unique Wiener filter representations,
which consequently prevents the unique recovery of a LCDN based on conditional Granger causality. 
A possible LCDN graph inferred under this omission is shown in Fig.~\ref{fig:topo} (b),
which contains 41 edges and exhibits markedly more complex internal connections than the corresponding LRDN. 
Moreover, in an extreme scenario where all off-diagonal entries of $W(z)$ are constant,
a complete LCDN graph with $132$ edges may be obtained---a result that would severely undermine subsequent prediction or control tasks.\\
Using least square methods, we obtain unbiased estimates of the Wiener filters $S(z)$ 
in Theorem~\ref{thm:WienerF} and $H(z)$ in \eqref{eq:ym=Hyl}.
The existence of all edges in $E$ of the LRDN graph is tested based on Theorem~\ref{thm:GrangerLDRN}.
We find that the reconstructed graph under the LRDN model from the sampled data perfectly matches the original graph in Fig.~\ref{fig:topo} (a).
The above results demonstrate that LRDN offers an efficient  and well-posed modeling framework for low rank processes, 
while the employed Wiener filters provide a reliable and consistent approach for network topology learning.
\section{Conclusions}\label{sec:con}
In this work we have addressed the modeling and topology estimation of linear causal dynamical networks exhibiting low rank output processes.
To this end, we have proposed the low rank dynamical network (LRDN) model, 
which provides a parsimonious and identifiable representation of the internal correlations.
The topology of an LRDN is recovered using causal Wiener filters: 
one capturing the deterministic relation between the subprocesses in the special feedback model, 
and the other operating under the constraint of strictly causal self-loops. 
We establish a fundamental necessary and sufficient condition that 
links the presence of an edge to conditional Granger causality, 
or equivalently, to the non-zero entries of the corresponding Wiener filters.
A simulation example demonstrates the parsimony of the LRDN framework and the consistent performance of our topology estimation method.
Further details and network identification under output errors will be discussed in a forthcoming journal version. 

\bibliography{ifacconf}             % bib file to produce the bibliography
                                                     % with bibtex (preferred)

%\appendix

\end{document}